\documentclass[letterpaper]{article} 
\usepackage{aaai24}  
\usepackage{times}  
\usepackage{helvet}  
\usepackage{courier}  
\usepackage[hyphens]{url}  
\usepackage{graphicx} 
\usepackage{natbib}  
\usepackage{caption} 
\usepackage{algorithm}
\usepackage{multirow}
\usepackage{amsmath}
\usepackage{amssymb}
\usepackage{booktabs}
\usepackage{algorithmicx}  
\usepackage{algpseudocode}

\usepackage{newfloat}
\usepackage{listings}
\DeclareCaptionStyle{ruled}{labelfont=normalfont,labelsep=colon,strut=off} 
\floatstyle{ruled}
\newfloat{listing}{tb}{lst}{}
\floatname{listing}{Listing}
\title{MEAOD: Model Extraction Attack against Object Detectors}
\author{
    Zeyu Li\textsuperscript{\rm 1}, Chenghui Shi\textsuperscript{\rm 1}, Yuwen Pu\textsuperscript{\rm 1}, Xuhong Zhang\textsuperscript{\rm 2}, Yu Li\textsuperscript{\rm 3},\\
    Jinbao Li\textsuperscript{\rm 4}, Shouling Ji\textsuperscript{\rm 1}
}
\affiliations{
    \textsuperscript{\rm 1}College of Computer Science and Technology, Zhejiang University\\
    \textsuperscript{\rm 2}School of Software Technology, Zhejiang University\\
    \textsuperscript{\rm 3}School of Computer Science and Technology, Harbin Institute of Technology\\
    \textsuperscript{\rm 4}School of Mathematics and Statistics, Qilu University of Technology
    
}

\usepackage{bibentry}

\begin{document}
\maketitle

\begin{abstract}
The widespread use of deep learning technology across various industries has made deep neural network models highly valuable and, as a result, attractive targets for potential attackers. Model extraction attacks, particularly query-based model extraction attacks, allow attackers to replicate a substitute model with comparable functionality to the victim model and present a significant threat to the confidentiality and security of MLaaS platforms. While many studies have explored threats of model extraction attacks against classification models in recent years, object detection models, which are more frequently used in real-world scenarios, have received less attention. In this paper, we investigate the challenges and feasibility of query-based model extraction attacks against object detection models and propose an effective attack method called MEAOD. It selects samples from the attacker-possessed dataset to construct an efficient query dataset using active learning and enhances the categories with insufficient objects. We additionally improve the extraction effectiveness by updating the annotations of the query dataset. According to our gray-box and black-box scenarios experiments, we achieve an extraction performance of over 70\% under the given condition of a 10k query budget.

\end{abstract}

\section{Introduction}

Over the past few years, neural network-based deep learning technologies have seen rapid development and adoption across various fields, including self-driving \cite{self-driving}, recommendation systems \cite{recommendation}, and facial recognition \cite{face-recognition}. However, training these models always requires vast amounts of data and computing power, making well-trained models valuable intellectual properties. As a result, such models have become targets for query-based model extraction attacks that threaten Machine Learning as a Service (MLaaS) platforms \cite{stealML}. Attackers can replicate the functionality of the victim model with less cost, presenting challenges to the secure application of MLaaS platforms.

Prior research on this topic has mainly focused on classification models \cite{CloudLeak},  or other fields such as graph neural network models \cite{gnn} and image encoders \cite{StolenEncoder}. The general workflow of the model extraction attack is as follows: Firstly, the attacker collects samples from publicly available datasets or generates images with generation models to construct an unlabeled dataset as a query dataset, which contains samples in the victim task scene. Then the attacker uses the samples in the query dataset to query the victim model and obtain corresponding outputs. Finally, the attacker uses the query dataset to train a substitute model with a designed extraction loss function.

Despite the model extraction attack methods mentioned above, they may not be applicable to object detection models. Object detection models \cite{YOLO, YOLOv5, focal, fast-rcnn}, widely used in areas such as self-driving and industrial inspection, are commonly deployed in deep learning cloud services and are targets for model extraction attacks. But to our knowledge,  only \emph{Imitated Detectors} \cite{ImitatedDetectors} focused on object detection models. It only requires a few real samples and uses a text-to-image model \cite{VQgan,minGPT, DALL-E} to generate objects and synthesize images. However, \emph{Imitated Detectors} may fail to generate high-quality objects, and the synthetic images in the query dataset lack backgrounds. These limitations reduce the information from the victim model and thus limit the performance of \emph{Imitated Detectors}, especially for real-world scenario datasets. Therefore, based on the current research status, we investigate the threat of model extraction attacks against object detection models under a data-sufficient setting and propose an efficient attack method.

We highlight the challenges posed by model extraction attacks against object detection models compared to classification models. 1) \emph{Low extraction efficiency}: existing model extraction attacks commonly employ techniques like active learning to construct the query dataset, thereby enhancing the efficiency of the extraction process. However, object detection models contain both detection and classification subtasks, which cause the existing methods designed primarily for classification models to suffer an extraction efficiency loss while transferring to the object detection field; 2) \emph{Performance limitations for rare categories}: model extraction attack on object models need a great demand for the quality of the query dataset. However, There is a significant imbalance of the number of objects between each category in the query dataset. The imbalance would affect the extraction result, causing the substitute model to perform poorly in categories with fewer objects; 3) \emph{Low-quality detection result from the victim model}: the output processing mechanisms of object detection models, such as Non-Maximum Suppression (NMS), prevent the model from generating predictions for objects outside the distribution. Besides, the object detection tasks are typically more complicated than the classification, and the object detection models generally have lower accuracy. These limitations make the victim model prone to returning low-quality results.

Targeting the challenges mentioned above, in this paper, we propose MEAOD, a novel model extraction method against object detection models in a data-sufficient setting. The method generates query datasets using active learning methods, which consider both classification and detection subtasks to overcome the challenge of low extraction efficiency. Additionally, to address the issue of limited performance in rare categories, we introduce a second stage in the query dataset construction and filter samples from the Internet to enhance the query dataset. To mitigate the impact of the low-quality detection result from the victim model on the performance of the substitution model, we adopt a dynamic update method for the query dataset labels. Using the proposed approach, an attacker can replicate the victim model with a few queries and publicly available data. In summary, the contributions of our work are as follows:

\begin{itemize}
    \item We investigate the feasibility and challenges of model extraction attacks against object detection models, and develop a model extraction attack method specifically for object detection models.
    
    \item We propose a query dataset construction framework, which contains a simple yet effective active-learning method considering both detection and classification, and a fitness-based dataset enhancement to select informative samples from the Internet.
    
    \item We propose a scale-consistency based annotation updating mechanism to promote the extraction performance by mitigating the incorrect results from the victim model.

    \item Our experiments on two task scenarios and four model architectures demonstrate the effectiveness of MEAOD.
\end{itemize}

\begin{figure*}
  \centering
  \includegraphics[width=5.5in]{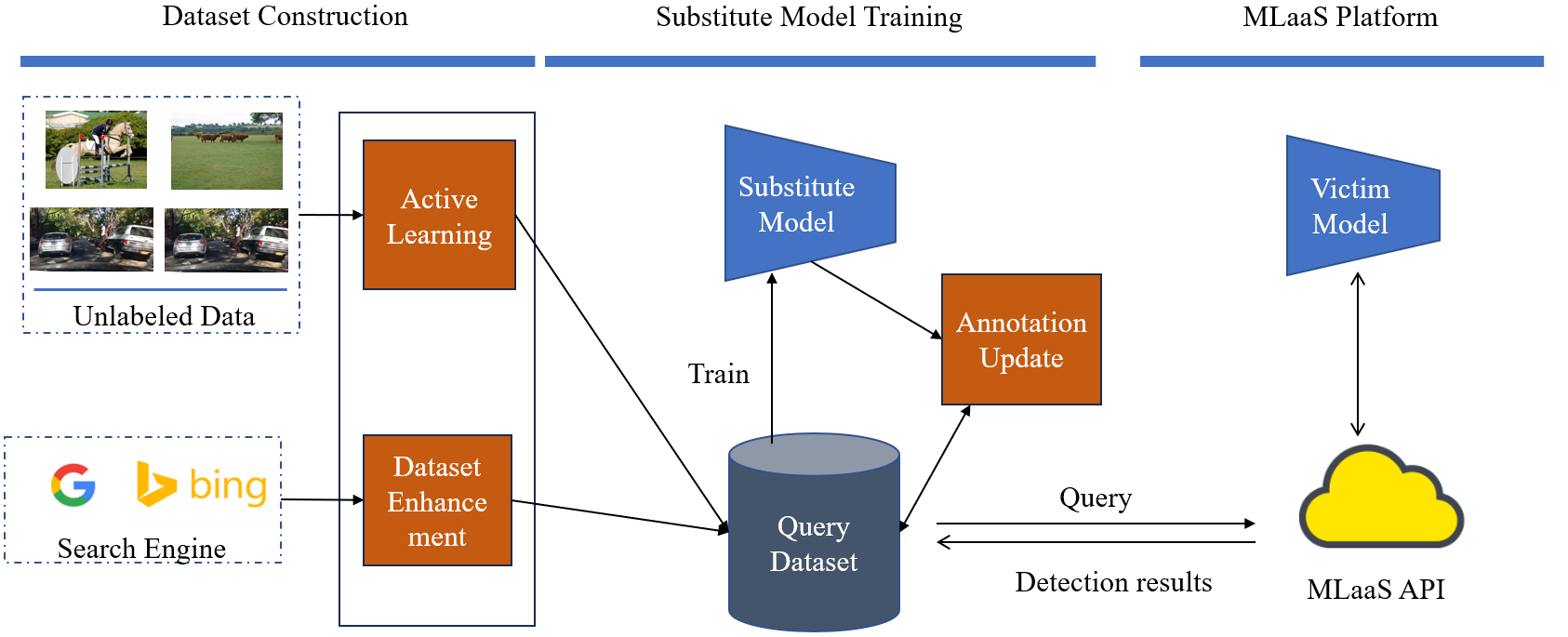}
  \caption{The workflow of MEAOD. Attackers first generate a query dataset using a two-stage method containing an active-learning based dataset construction and a dataset enhancement for rare categories. After querying the victim model, the results need to be filtered by confidence and processed through an update mechanism before being used for substitute model training.}
  \label{fig:1}
\end{figure*}


\section{Related Work}

\subsection{Model extraction attack}
Currently, the existing query-based model extraction attacks can be split into data-sufficient setting \cite{CloudLeak, KnockoffNets, PRADA}, and data-limited setting \cite{Dast, MAZE, ESattack} based on the number of samples available to the attacker. In the data-sufficient scenario, the attacker typically collects an unlabeled dataset of the target task. CloudLeak \cite{CloudLeak} utilized active learning to construct an efficient query dataset and effectively fitted the classification boundaries of the victim model with adversarial examples. 
In contrast, in the data-limited scenario, the attacker has only a few or even no real samples and must synthesize samples using generation models like GAN \cite{GAN}. MAZE \cite{MAZE} used GAN-generated samples to construct the query dataset and obtained approximate victim model gradients by zero-order gradient estimation to optimize the training of generators. The further work \cite{data-free&hard-label} explored data-free model extraction in a hard-label setting. Researchers have also started to explore extracting other kinds of models, such as encoders  \cite{StolenEncoder, Cont-Steal}, GNN \cite{gnn,gnn:taxonomy,advgnn,gnnlinks}, generation models \cite{stealgan,artsteal} and large language models  \cite{StealLanguage}. However, research against object detection models is still rare.

\subsection{Object detection}

The object detection task is a classic machine-learning problem that aims to identify and locate objects of a specific class in a given image by labeling their positions with a bounding box. Current popular object detection models can be categorized into one-stage and two-stage detectors. Two-stage detectors divide the object detection task into object localization and classification tasks and use two separate parts to complete the task, with representative models such as RCNN \cite{rcnn} and Fast-RCNN \cite{fast-rcnn}. In contrast, one-stage detectors represented by the YOLO series models \cite{YOLO}, SSD \cite{SSD}, and RetinaNet \cite{retinanet} merge the two subtasks into a network, leading to faster prediction and better support for real-time applications. Both types of these methods are anchor-based, where anchor-related hyperparameters influence the performance of the models and different datasets require the redesign of anchors. So anchor-free object detection methods \cite{CornerNet,fcos, ExtremeNet, AOPG} that can overcome these limitations have received widespread attention in recent years. 

\section{Threat model}

\subsection{Adversary's goal}
 A model extraction attack against the object detection model should achieve such goals: a) the attacker can acquire a substitute model that has similar functionality to the victim model; b) the attacker must spend less time and computational power than the training cost of the victim model; c) the attack should not require excessive query operations to avoid detection by the victim model's owner.
\subsection{Adversary's ability}

We assume that an attacker can access the victim model through a black-box API and obtain detection results, containing the detection bounding box, the object category, and a score indicating the confidence of the detection box. In object detection tasks, attackers are prone to collect task-related data, so we use a data-sufficient setting. In this setting, attackers can collect task-related data through the Internet or publicly available datasets, thereby obtaining an unlabeled dataset that approximates the distribution of the victim dataset. Additionally, the attacker can gather details about the different types of objects the victim model can identify. They might achieve this by studying documents or interacting with the victim model using samples that encompass potential categories. For the architecture knowledge of the victim model, we divide the discussion into two scenarios: gray-box and black-box scenarios. In the gray-box scenario, the attacker can access to the architecture information of the victim model, which enables the attacker to expedite the attack process by using a structure and setting similar to the victim model, while the attacker in the black-box scenario lacks such knowledge.

\section{Method}

MEAOD is divided into two main parts: the construction of the query dataset $D_q$ and the training of the substitute model $M_s$. Firstly, we propose using a two-stage dataset construction approach to generate an efficient and generalized query dataset. Secondly, we propose an annotation update mechanism based on dynamic confidence threshold and scale-consistency to mitigate the impact of lower-quality detections returned by the victim model. In the following sections, we will explain in detail the design of MEAOD, the framework of which is presented in Figure \ref{fig:1}.

\begin{algorithm}[h]  
\small
    \caption{The two-stage dataset construction method generates a efficient query dataset $D_q$ for models extraction using attacker's unlabeled dataset $D_I$}  
    \begin{algorithmic}[1] 
            \Require Victim model $M_v$, Initial weight of substitute model $\theta_s$,Initial dataset collected by attacker $D_I$, Number of samples selected in each iteration $k_a$, Number of samples selected in the second stage $k_s$, Number of activate learning iterations
            \Ensure Query dataset $D_q$
            \Procedure{QueryDataset}{$M_v,\theta_s,D_I,k_a,k_s,$}  
                \State $D_q\gets \{\}$
                \State $M_s\gets initial\quad weight\quad by \quad \theta_s$
                
                \For{$i = 0 \to I-1$}
                    \For{each $x \quad\in \quad D_I \backslash D_q$}
                        \State Calculate and record $U(x)$
                    \EndFor 
                    \State Select $k_a$ samples $D_a$ in $D_I \backslash D_q$ with highest $U(x)$
                    \State Query $M_v$ with $D_a$
                    \State $D_q \gets D_q \cup D_a$
                    \State Train $M_s$ with $D_q$
                \EndFor   
                \State Analyze class distribution in $D_q$ and recognize $C_r$
                \State Collect images of $C_r$ via search engine as $D_s$
                \For{each $x \quad\in \quad D_s$}
                        \State Calculate and record $S(x)$
                    \EndFor 
                \State Select $top-k_s$ samples $D_s$ in $D_s$ with highest $S(x)$
                \State Query $M_v$ with $D_s$
                \State $D_q \gets D_q \cup D_a$
                
                \State \Return{$D_q$}  
            \EndProcedure  
    \end{algorithmic}
\end{algorithm}  
\subsection{Two-stage query dataset construction}
To make the attack more efficient and balance different categories, we introduce an active-learning based dataset construction and a dataset enhancement as our two-stage dataset construction. In the first stage, we introduce an uncertainty-based active learning method to reduce the query budget. In the second stage, we propose a fitness-based filtering method to select informative samples from the Internet to enhance the extraction performance for rare categories. Algorithm 1 shows the complete two-stage attack algorithm.

\subsubsection{First stage: active learning based dataset construction}

In this stage, the attacker constructs a query dataset by selecting samples from the unlabeled dataset through an active learning mechanism. Active learning \cite{activesurvcy} is a commonly used technique to accelerate the training of deep learning models. Among the model extraction attacks for classification models \cite{CloudLeak,activethief}, active learning-based methods are widely used to improve attack efficiency and reduce query costs. Basically, the active learning process involves three steps: inputting samples into the substitute model $M_s$, calculating the value score for the samples by using the model output,  and selecting the samples with higher value scores to form the query dataset. In our attack, we use \emph{uncertainty} as the metric to assess the potential value of a sample in querying the victim model. We take an iterative approach, expanding $D_q$ and training $M_s$ orderly at each step.

The uncertainty of the model for a given sample is divided into classification uncertainty and localization uncertainty, corresponding to the two subtasks of object detection. For a given sample $x$, we calculate the uncertainty based on $M_S(x)$. In terms of classification uncertainty, we refer to \cite{ODactive-learming} and calculate the uncertainty using the two highest values in the classification confidence. The closer these values are, the higher the uncertainty, indicating that the substitute model cannot give deterministic classification results. Specifically, the classification uncertainty for the $i$th detection box $o_i$ for sample $x$ is specifically calculated as:
\begin{equation}
U_c(o_i) = C_{obj}(o_i) \cdot (1-(\max \limits_{c_1 \in K}p(c_1|o_i)-\max \limits_{c_2 \in K \backslash c_1}p(c_2|o_i)))^2
\end{equation}
where $C_{obj}(o_i)$ is the objectness confidence of $o_i$ and K is all categories of the target task. In terms of localization uncertainty, we select all prediction bounding boxes $b_{i,j}$ of the substitute model for the $i$th detection target $o_i$ in the sample, select the bounding box with the largest confidence $b_{i,0}$, and calculate the interaction ratio between other $b_{i,j}$ and $b_{i,0}$. A lower interaction ratio means higher uncertainty of the substitute model for localizing $o_i$, indicating that $o_i$ differs more from the samples in the existing query dataset. Specifically, the localization uncertainty is calculated as follows:
\begin{equation}
U_p(o_i) = \displaystyle \sum_{j=1}^{n} (1-iou(b_{i,0},b_{i,j}))
\end{equation}
where $iou()$ is the interaction ratio of the two labeled boxes, and the overall uncertainty of the $M_s$ for sample $x$ is:

\begin{equation}
U(x) = \displaystyle \sum _{o_i}U_c(o_i)\cdot U_p(o_i) 
\end{equation}

Based on the two uncertainty metrics above, we select the highest-rated samples to add to the query dataset and query the victim model, using its returned detection as labels of the query dataset. Our active-learning method is simple yet effective. It can be replaced by the state-of-art methods designed for object detection \cite{activeOD}, but this may cost more computation source for attackers.

\subsubsection{Second stage: Internet sample dataset enhancement}

The initial unlabeled dataset $D_I$ often differs from the victim model dataset in various aspects, with the most significant being that objects of specific categories are relatively rare or even missing in the attacker-owned dataset. For instance, a victim model focused on self-driving may collect uncommon foreground objects, such as traffic cones and signs, in its dataset. In contrast, these objects are likely to be missing in the dataset collected by an attacker, mainly containing pedestrians and vehicles. 
The details of the imbalance between different categories are presented in the supplementary material. We refer to the categories with fewer objects in the query dataset as $C_r$, while the other categories as $C_f$. In the second stage, we augment the query dataset for these rare categories with images from the Internet.

Since the attacker knows all the categories of the victim model, the missing categories can be identified by counting the number of annotations of different categories in the query dataset. We generate the search keyword based on the target task scenario and the name of the missing category to retrieve relevant image samples from search engines like Google. However, there is a significant amount of noise in samples from the Internet, which is considered irrelevant to our attack. So we use the substitute model after the first stage of dataset construction to select the in-distribution data in the collected Internet samples. Additionally, we also consider the image size as an important factor, since larger images tend to contain more information. Therefore, we design a validation criterion to check the suitability of the collected samples to be added to the query dataset:
\begin{equation}
S(x)= \alpha*mean(Conf(x))+ std(Conf(x))\cdot size(x)
\end{equation}
where $Conf(x)$ is the objectness confidence of $M_s$ for all bounding boxes of sample $x$, and $size(x)$ is the size of sample $x$. $mean$ and $std$, respectively, denote the mean and standard deviation of the distribution. In this scoring process, the mean value reflects the degree of fitness between $x$ and the target task scenario, while the standard deviation indicates whether $x$ contains high-quality objects significantly different from the background. We use the batch of samples with the highest score to enhance the query dataset and eliminate the noise with a lower score.

\begin{figure}
  \centering
  \includegraphics[width=3in]{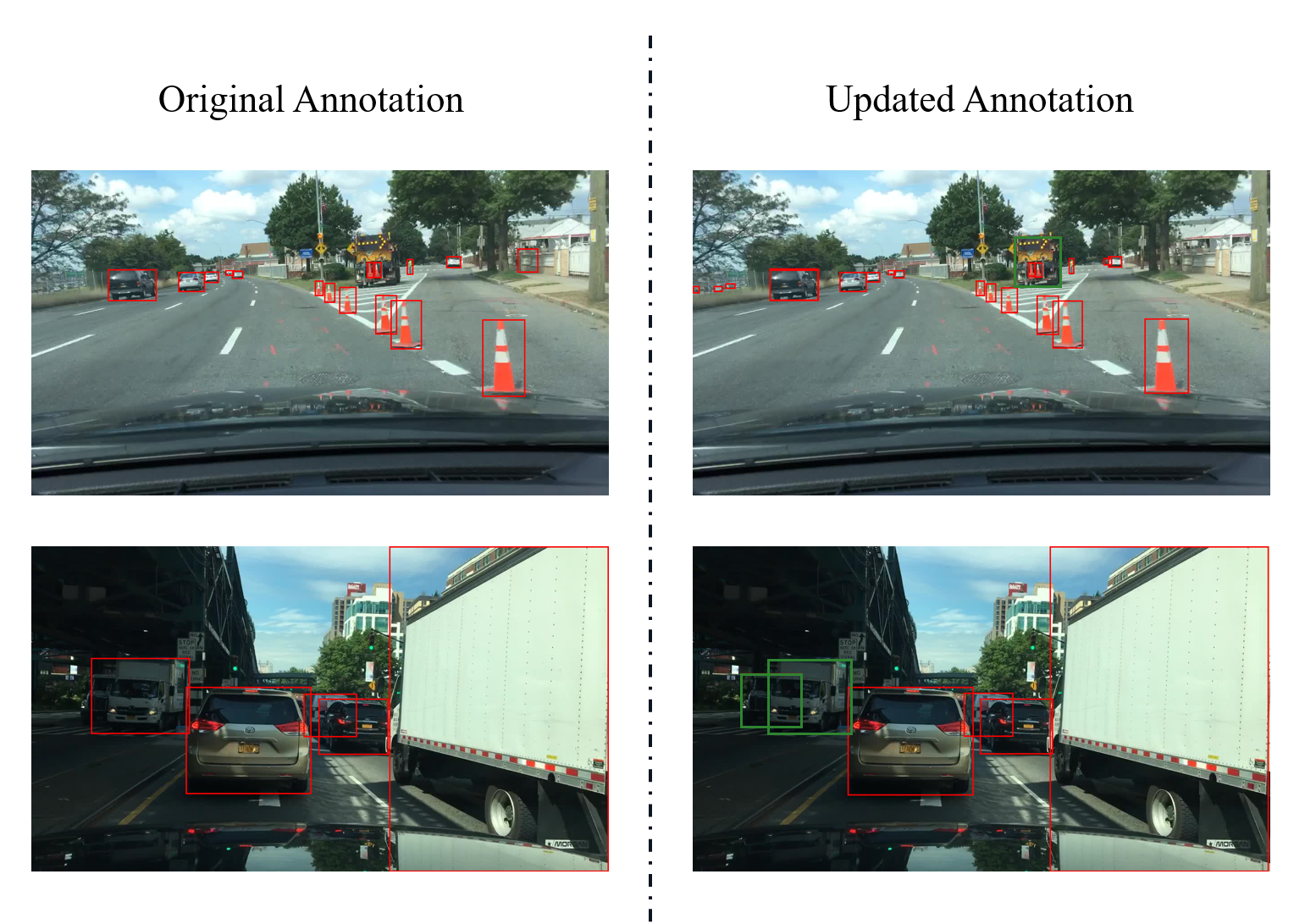}
  \caption{The confidence filter and scale-consistency based annotation update mechanism mitigate the mistake made by the victim model. We highlight the annotations updated in the mechanism. Left: the original detection results returned by the victim model. Right: annotations after the process.}
  \label{fig:2}
\end{figure}

\subsection{Substitute model training}
In existing model extraction methods against other models, an attacker can design a distillation loss function that computes the difference between the query results of the local substitute model and the victim model. However, the anchor-based object detection model is not a strict end-to-end model, and its model output is converted into detection results only after NMS, so the attacker can only use the inherent loss function of the model architecture to train the substitute model. In this setting, the training of the substitute model $M_s$ is very sensitive to the quality of the returned annotations of the victim model, and the presence of low-quality annotations can significantly affect the extraction result. Two main types of low-quality annotations are likely to occur in the returned information of the victim model: 
\begin{itemize}
    \item redundant annotations: labeling the background as the foreground object;
    \item missing annotations: failing to detect foreground objects.
\end{itemize}

Among these two types, the issue of redundant annotations can be effectively mitigated by implementing a dynamic confidence threshold after the query. When receiving annotations from the victim model, annotations with confidence below a dynamic threshold $theta_c$ are rejected. This thresh
old is established separately for each category and increases with the number of detection in the query results for that category. Furthermore, we tackle the problem of missing annotations by introducing an annotation update mechanism. In the object detection task, models are expected to be scale-consistent and capable of detecting objects of different sizes. Detections that retain stable localization and classification across different sizes generally exhibit higher confidence. Therefore, detections with scale consistency are more likely to be correctly labeled for objects in the same sample of different scales. Motivated by this intuition, we propose a dynamic annotation update method based on scale consistency. This approach involves using substitute models to detect the same sample in different sizes, and objects showing scale consistency are incorporated into the ground truth of the query dataset. The metric for determining the scale consistency of the detection result $o$ are as follows:
\begin{equation}
C(o) = \displaystyle \min_{1<=i<=n}\displaystyle \max_{1<=j<=m_i}iou(o,o_{i}^{j})
\end{equation}
where $n$ is the total number of scales and $o_{i}^{j}$ represents jth bounding box under the ith scale. In the calculation, we select all detection boxes under a scale as candidates, calculate the scale consistency of each detection, and add it to the label if it is greater than the threshold $\theta _u$. We show several examples with processed annotations in Figure \ref{fig:2}.

\begin{table}[h]
\small
  \centering
  \caption[Short Caption]{The victim model performance on the nuImage and COCO dataset}
  \renewcommand{\arraystretch}{1.1}
  \begin{tabular}{@{}lcccc@{}}
    \toprule
    dataset & YOLOv3 & YOLOv5 & YOLOv7 & Fast-RCNN \\

    \hline
    nuImage & 0.742 & 0.719 & 0.794 &0.615 \\
    COCO & 0.653 & 0.641 & 0.697 & 0.554 \\
    \hline
    \bottomrule
  \end{tabular}
  
  \label{tab:victim}
\end{table}

\begin{table*}[h]
\small
  \centering
  \caption{Gray-box object detection model extraction evaluation}
  \renewcommand{\arraystretch}{1.2}
  \begin{tabular}{@{}l|cc|cc|cc|cc@{}}
    \toprule
    \multirow{3}{*}{Method} & \multicolumn{4}{c|}{COCO} &  \multicolumn{4}{c}{nuImage}\\
    \cline{2-9}
    & \multicolumn{2}{c|}{YOLOv5} & \multicolumn{2}{c|}{YOLOv7} & \multicolumn{2}{c|}{YOLOv5} & \multicolumn{2}{c}{YOLOv7} \\
    \cline{2-9}
    & map50 & map50:95 & map50 & map50:95 & map50 & map50:95 & map50 & map50:95 \\
    \hline
    random &0.399(62.2\%) &0.239 & 0.382(54.8\%)&0.233 & 0.505(70.2\%)&0.305 &0.439(55.3\%) &0.297 \\
    Cloudleak & 0.391(61.0\%)&0.233 & 0.39(55.9\%)& 0.192& 0.551(76.6\%)&0.345 & 0.480(60.5\%)&0.321 \\
    Ours &\textbf{0.457(71.3\%)} &\textbf{0.291} &\textbf{0.455(65.3\%)} & \textbf{0.302}& \textbf{0.579(80.5\%)}&\textbf{0.354} &\textbf{0.554(69.7\%)} &\textbf{0.389}\\
    \hline
    \bottomrule
  \end{tabular}
  
  \label{tab:grey}
\end{table*}

\begin{table*}[h]
\small
  \centering
  \caption{Black-box object detection model extraction evaluation}
  \renewcommand{\arraystretch}{1.1}
  \begin{tabular}{@{}l|c|ccc|ccc@{}}
  
    \toprule
    \multicolumn{2}{c}{\multirow{2}*{method}} & \multicolumn{3}{|c|}{COCO} &  \multicolumn{3}{c}{nuImage}\\
    \cline{3-8}
    \multicolumn{2}{c}{}& \multicolumn{1}{|c}{random} & \multicolumn{1}{c}{Cloudleak} & \multicolumn{1}{c|}{Ours} & \multicolumn{1}{c}{random} & \multicolumn{1}{c}{Cloudleak} & \multicolumn{1}{c}{Ours} \\
    \hline
    \multirow{2}{*}{YOLOv5/YOLOv7} & map50 &0.388(55.7\%) &0.399(57.2\%) &\textbf{0.454(65.1\%)} &0.486(61.2\%) &0.494(62.2\%) &\textbf{0.576(72.5\%)}\\
    \cline{2-2}
    & map50:95&0.229 &0.235 &\textbf{0.285}  &0.298 &0.3 &\textbf{0.345}\\
    \hline
    \multirow{2}{*}{YOLOv5/YOLOv3} & map50 &0.377(57.7\%) &0.391(59.9\%) &\textbf{0.424(64.9\%)} &0.481(64.8\%) &0.472(63.6\%) &\textbf{0.552(74.4\%)}\\
    \cline{2-2}
    & map50:95&0.208 &0.318 & \textbf{0.258}& 0.287&0.284&\textbf{0.331}\\
    \hline
    
    \multirow{2}{*}{YOLOv5/Fast-RCNN} & map50 &0.4(72.2\%) &0.397(71.7\%) &\textbf{0.447(80.7\%)} &0.412(67.0\%) &0.424(68.9\%) &\textbf{0.523(85.0\%)}\\
    \cline{2-2}
    & map50:95&0.222 & 0.22&\textbf{0.245} & 0.212& 0.223&\textbf{0.281}\\
    \hline
    \multirow{2}{*}{YOLOv7/YOLOv5} & map50 & 0.401(62.6\%)&0.382(59.6\%)&\textbf{0.435(67.9\%)} &0.439(61.1\%) &0.447(62.2\%) &\textbf{0.543(75.5\%)}\\
    \cline{2-2}
    & map50:95& 0.256&0.239 &\textbf{0.28} &0.284 &0.286 &\textbf{0.341} \\
    \hline
    \bottomrule
  \end{tabular}
  
  \label{tab:black}
\end{table*}

\section{Experiment}

According to the goals of MEAOD, we design experiments to evaluate attack effectiveness and cost. In this section, we first present our experimental setup, followed by experiments on attack effectiveness and cost, as well as ablation experiments. Finally, we discuss the experiment results and several specific study cases.
\subsection{Experiment setting}

In this subsection, we introduce our experimental setup from three aspects: dataset, model architecture, and metrics. Extra information is presented in the supplementary material.
\subsubsection{Dataset}
In this paper, we mainly consider general object detection and autonomous driving object detection scenarios and use four widely used datasets as follows.
\begin{itemize}
    \item \textbf{COCO dataset} \cite{COCO}: the most commonly used generic object detection dataset, containing 80 categories. Like the setting in  \cite{ImitatedDetectors}, we only extract the 20 categories from the VOC dataset.
    \item \textbf{VOC dataset} \cite{VOC}: a commonly used generic object detection dataset that contains fewer samples than the COCO dataset, with only 20 categories. 
    \item \textbf{nuImages dataset} \cite{nuImages}: a self-driving scenario dataset released by Montional, which originally has 23 categories, and we integrate some similar categories into ten categories according to another dataset nuScenes \cite{nuScenes} from the same company. 
    \item \textbf{BDD100k dataset} \cite{bdd100k}: a large-scale public self-driving dataset released by Berkeley University AI Lab, containing 100,000 samples. 
    
\end{itemize}

As the VOC dataset and BDD100k dataset have fewer categories and lower image quality, We use the training set of them as attacker-owned unlabeled datasets, while the COCO dataset and nuImage dataset as the victim dataset to train the victim model and test the substitute model.

\subsubsection{Architecture}
Regarding the model architecture of the victim model, we use the YOLOv3 \cite{YOLOv3}, YOLOv5 \cite{YOLOv5}, and YOLOv7 \cite{YOLOv7} model structures from the YOLO series for one-stage detectors, and mainly used Fast R-CNN \cite{fast-rcnn} for two-stage detectors. All four object detection architectures are used for the victim model, while the substitute model only uses three one-stage model architectures from the YOLO series.
\subsubsection{Metric}
The most common performance metric for object detection tasks is mAP, calculated by computing the area under the precision-recall curve for each class and then taking the average value. We mainly use mAP@50 and mAP@50:95 to measure model performance. Additionally, we use the relative proportion of the substitute model's mAP@50 to the victim model's mAP@50 as a metric to measure the effectiveness of the model extraction attack.

\subsection{Extraction attack evaluation}
In this subsection, we evaluate the performance of our model extraction attack and compare it with existing methods. The existing attack method for object detection models, Imitated detector \cite{ImitatedDetectors}, is not open-sourced and requires an additional text-to-image model, so we migrate the extraction attack method initially designed for classification models to the object detection domain for comparison. In this subsection, we mainly migrate Cloudleak \cite{CloudLeak}, an existing model extraction attack method for data-sufficient settings, to the object detection task domain. Since adversarial examples used in Cloudleak would affect the training of substitute models, we use clean samples to compose the query dataset. In addition, we also use the method of randomly selecting the data possessed by the attacker as the query dataset as a baseline. We conduct experiments in both gray-box and black-box scenarios. In the gray-box scenario, the attacker uses the same model architecture as the victim model. In the black-box scenario, we used multiple model architectures to verify the performance of the model extraction attack. 

The attack results in the gray-box scenario are shown in Table \ref{tab:grey}. MEAOD can extract more than 70\% of the performance of the victim model when the query dataset contains 10,000 samples, with a sample ratio 3:2 between the two stages of dataset construction, and can achieve better attack effects compared to the other two baselines. In the black-box scenario, the attack's results are shown in Table \ref{tab:black}. The substitute model can also achieve a relative accuracy of over 70\%. Despite the attacker's inability to use the same model structure as the victim model, the attack effectiveness is only slightly reduced. This observation highlights that, in a model extraction attack against the object detection model, the crucial determinant of success or failure lies in the attacker's ability to gather high-quality query data. This factor holds greater significance compared to the attacker's knowledge of the victim model architecture.

\subsection{Attack cost}
In this subsection, we quantitatively compare the costs of training the victim model and the attacker's substitute model in terms of training time and computational power. All the experiments in this subsection are conducted on 2 RTX3090ti GPUs. As shown in Table \ref{tab:cost},  compared to the victim model trained on the original dataset, the cost of data collection is within a controllable range, and the substitute model requires less training time because only a small portion of the unlabeled dataset $D_I$ is involved in the training.

\begin{table}[h]
\small
  \centering
  \caption{The attack cost of MEAOD}
  \renewcommand{\arraystretch}{1.1}
  \begin{tabular}{@{}l|cc|cc@{}}
    \toprule
    \multirow{2}{*}{Model} & \multicolumn{2}{c|}{nuImage origin} &  \multicolumn{2}{c}{nuImage extraction}\\
    \cline{2-5} 
    & data & time (h) & data & time (h) \\
    \hline
    YOLOv5 & 60655& 80.52& 93690&27.81 \\
    YOLOv7 & 60655& 99.31& 93690&  26.88 \\
    YOLOv3 &  60655& 87.65& 93690& 28.13 \\
    Fast-RCNN & 60655& 126.5& 93690& 27.56\\
    \hline
    \bottomrule
  \end{tabular}
  
  \label{tab:cost}
\end{table}
\subsection{Ablation study: effectiveness of each module}
In this subsection, we conduct an ablation study to evaluate the effectiveness of different modules in our proposed method with the same query budget setting as subsections above. We divide MEAOD into three parts: A) constructing the query dataset with active learning, B) enhancing the query dataset with Internet samples, and C) updating the query dataset labels based on scale consistency. Based on these three parts, we designed the following ablation studies: (1) constructing the query dataset by random selection, denoted as [Ø], (2) constructing the query dataset with active learning, denoted as [A], (3) constructing the query dataset with active learning and dataset enhancement, denoted as [A, B], (4) constructing the dataset by random selection and enhancing it with Internet samples, denoted as [B], (5) constructing the query dataset with active learning and dynamically updating labels, denoted as [A, C]. Additionally, we compare these results with the experiments that enable all modules in the previous sections, denoted as [A, B, C]. All experiments are conducted using the YOLOv5 architecture. The experimental results, as shown in Table \ref{tab:ablation}, demonstrate the effectiveness of active learning and Internet data set enhancement by comparing [A] to [Ø] and [B] to [Ø]. Moreover, the comparison of [A, B, C] and [A, B] effectively verifies that the substitute model based on scale consistency can accurately update the annotation of the query dataset.

\begin{table}
\small
  \centering
  \caption{The ablation study on the effectiveness of each module in MEAOD}
  \begin{tabular}{@{}l|c|c|c|c@{}}
    \toprule
    \multirow{2}{*}{Method} & \multicolumn{2}{c}{COCO} &  \multicolumn{2}{c}{nuImages}\\
    \cline{2-5}
    & map50 & map50:95 & map50 & map50:95  \\
    \hline
    $\emptyset$ &0.399(62.2\%) &0.239 &0.505(70.2\%) &0.305 \\
    A &0.428(66.8\%) &0.247 &0.529(73.6\%) &0.317 \\
    B & 0.396(61.8\%)&0.242 & 0.495(68.8\%)&0.299  \\
    AB &0.428(66.8\%) &0.26 & 0.574(79.8\%)&0.351  \\
    AC &0.43(67.1\%) &0.25 & 0.552(80.5\%)&0.339 \\
    ABC &\textbf{0.457(71.3\%)} &\textbf{0.291} &\textbf{0.579(80.5\%)} & \textbf{0.354}\\
    \hline
    \bottomrule
  \end{tabular}

  \label{tab:ablation}
\end{table}

\subsection{Sensitive study: impact of different query budget}

\begin{figure}
  \centering
  \includegraphics[width=3.3in]{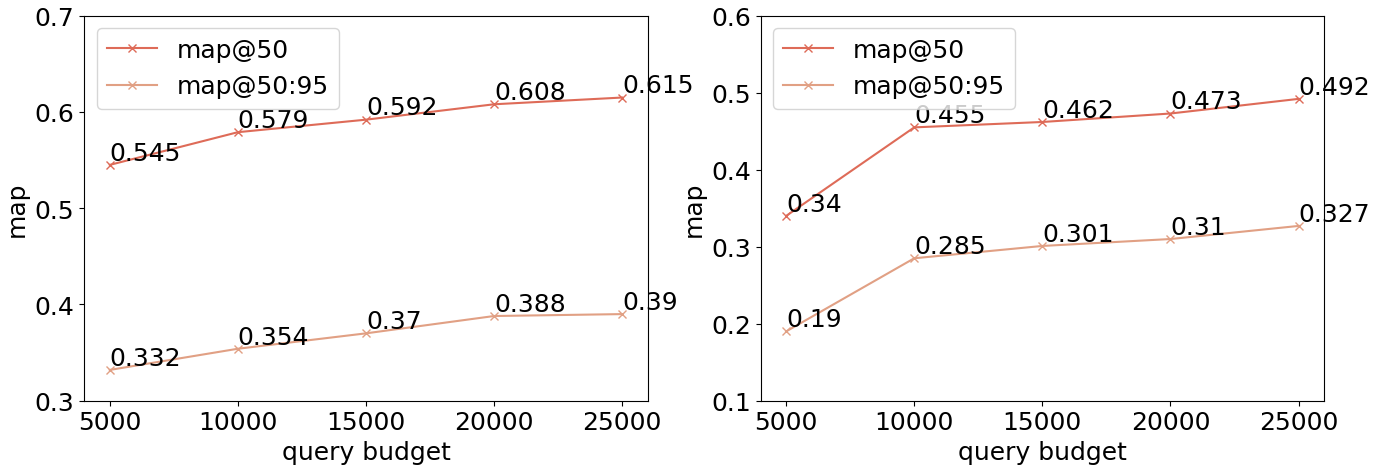}
  \caption{The sensitive study on the impact of query budget in MEAOD. Left:self-driving scenario with nuImage dataset. Right: general detection scenario with COCO dataset.}
  \label{fig:3}
\end{figure}

The query budget can significantly impact the extraction performance in model extraction attacks. This subsection investigates the effectiveness of MEOAD across different query budgets. We conduct experiments on YOLOv5  \cite{YOLOv5} model in the self-driving scenario and general detection scenario, analyzing the extraction performance achieved with an increased query budget. Our experiments are conducted across a range of query total frequencies, including 5k, 10k, 15k, 20k, and 25k, with a sample ratio of 3:2 between the two stages of dataset construction. As shown in Figure \ref{fig:3}, The experimental results indicate that the extraction performance increase as the query budget increases. Furthermore, as the query dataset reaches a larger proportion of the attacker's collected data, the extraction performance continues to converge, reaching about 86\% at most.

\subsection{Discussion}
\subsubsection{Partial attack}

\begin{figure}
  \centering
  \includegraphics[width=3.3in]{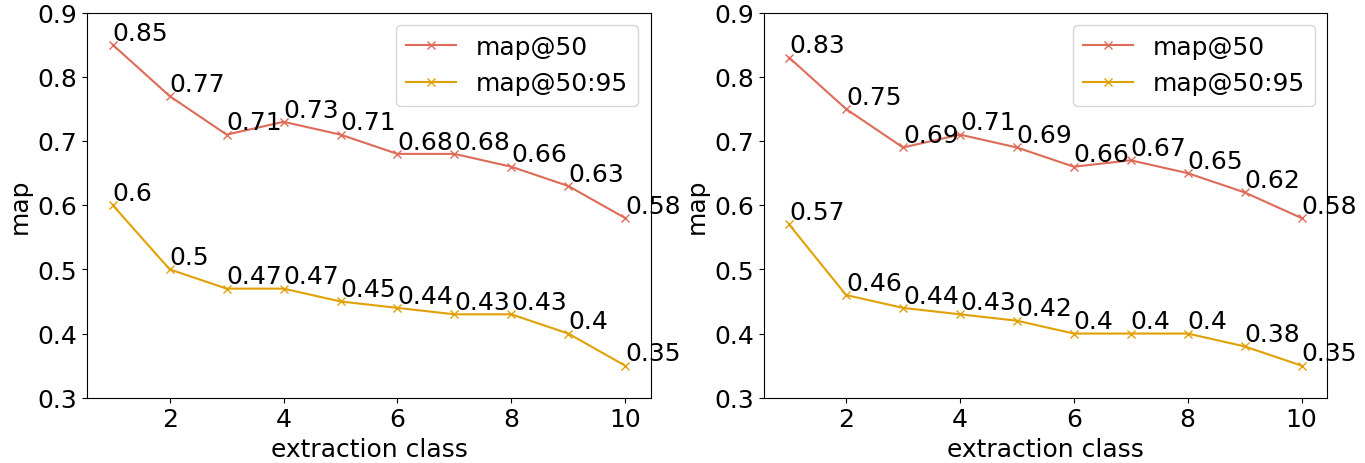}
  \caption{Extraction result of the partial attack. Left: YOLOv5 victim model. Right: YOLOv7 victim model.}
  \label{fig:4}
\end{figure}

In model extraction attacks against object detection models, the victim model may have many classes the attacker is not interested in. So, in this case, the attacker is more inclined to steal only a subset of all categories. For instance, in a self-driving scene, the attacker may be more interested in detecting frequently occurring objects like pedestrians, than other objects, like animals. We sort the categories in the nuImage dataset according to their frequency of occurrence and extract the top-n most frequently occurring categories. As shown in Figure \ref{fig:4}, the substitute model using the YOLOv5 architecture achieves better accuracy on the YOLOv5 and YOLOv7 victim models when only some essential categories are targeted. Therefore, Attackers can ignore the inessential categories to improve the extraction performance and reduce the attack cost.

\subsubsection{Possible defense}
Currently, the commonly used methods for defending model extraction attacks include model watermarking and fingerprinting. While in MEAOD, substitute models are trained using real samples and the original loss function and can't learn sufficient information about the victim model, such as decision boundary. So the watermark and fingerprint methods are ineffective in defending against MEOAD. A possible defense method is to confuse the confidence of the detection results by randomly modifying some high-confidence detection results to low confidence. When an attacker queries the victim model, they will automatically filter out these objects, resulting in the loss of the attack result. However, this may also affect the regular use of users.
\subsubsection{Limitation}
As mentioned above, MEAOD still requires the attacker to have a certain number of real samples to initialize the query dataset before the attack, which increases the cost and difficulty of extraction attacks in some highly confidential scenarios. Addressing this limitation poses a considerable challenge since attackers are unable to use out-of-distribution data to query the victim model, a strategy used in certain existing studies for classification models. This is due to the fundamental characteristic of object detection models, which typically do not provide any response to out-of-distribution data. Furthermore, the iterative training approach used for the substitute model might lead to overfitting of the samples incorporated into the query dataset during earlier stages. Such overfitting could potentially impact the overall performance of the model. We intend to explore potential solutions for these limitations in our future work.


\section{Conclusion}

In this paper, we explore the feasibility and challenges of model extraction attacks on object detection models and propose a two-stage query dataset construction method based on active learning. Additionally, to prevent the impact of lower-quality annotations returned by the victim model on the substitute model, we propose an improved method for dynamically updating dataset labels. Based on these
methods, we conduct experiments in two scenarios, revealing the vulnerability of object detection models based on cloud service APIs regarding model property rights.

\bibliography{aaai24}

\end{document}